\documentclass{article}
\usepackage{amssymb}
\usepackage{graphicx}
\usepackage{amsmath}

\newtheorem{theorem}{Theorem} [section]

\newtheorem{corollary}[theorem]{Corollary}

\newtheorem{example}[theorem]{Example}

\newtheorem{lemma}[theorem]{Lemma}

\newtheorem{proposition}[theorem]{Proposition}

\newenvironment{proof}[1][Proof]{\textbf{#1.} }{\ \rule{0.5em}{0.5em}}

\begin{document}

\author{Ana Meca$^{\dag}$\thanks{
Operations Research Center. Universidad Miguel Hern\'{a}ndez, Edificio
Torretamarit. Avda. de la Universidad s.n. 03202 Elche (Alicante), Spain}%
\thanks{%
Corresponding author. e-mail: ana.meca@umh.es} , Luis A. Guardiola$^{\dag }$
\ and Andr\'{e}s Toledo\thanks{%
Dpto. de Estudios de COHISPANIA, Compa\~{n}\'{\i}a Hispania de Tasaciones y
Valoraciones S.A. C. Portillo del Pardo, 14. 28023 Madrid, Spain}}
\title{p-additive games: a class of totally balanced games arising from
inventory situations with temporary discounts\thanks{
This work was partially supported by the Spanish Ministry of Education and
Science and Generalitat Valenciana (grants MTM2005-09184-C02-02,
ACOMP06/040, CSD2006-00032). Authors acknowledge valuable comments made by
the Editor and the referee.}}
\maketitle

\begin{abstract}
We introduce a new class of totally balanced cooperative TU games, namely $p$%
-additive games. It is inspired by the class of inventory games that arises
from inventory situations with temporary discounts (Toledo, 2002) and
contains the class of inventory cost games (Meca et al. 2003). It is shown
that every $p$-additive game and its corresponding subgames have a nonempty
core. We also focus on studying the character concave or convex and monotone
of $p$-additive games. In addition, the modified SOC-rule is proposed as a
solution for $p$-additive games. This solution is suitable for $p$-additive
games since it is a core-allocation which can be reached through a
population monotonic allocation scheme. Moreover, two characterizations of
the modified SOC-rule are provided.

\textbf{Key words:} $p$-additive games, inventory situations with temporary
discounts, totally balanced cooperative TU games, modified SOC-rule,
core-allocations.

\textbf{2000 AMS Subject classification:} 91A12, 90B05
\end{abstract}

\newpage

\section{Introduction}

In this paper we introduce the class of $p$-additive games. It is an
extension of the classes of inventory games with non discriminatory
temporary discounts as well as the class of inventory cost games. It turns
out to be a new class of totally balanced games with nice properties.

Inventory cost games are introduced and studied in Meca et al. (2004). In an
inventory cost game, a group of firms dealing with the ordering and holding
of a certain commodity decide to cooperate and make their orders jointly. To
coordinate the ordering policy of the firms, some revelation of information
is needed: the amount of revealed information between the firms is kept as
low as possible since they may be competitors on the consumer market. For
this class of games, Meca et al. (2004) focus on proportional division
mechanisms to share the joint cost. They introduce and characterize the
SOC-rule (Share the Ordering Costs) as an allocation rule for inventory cost
games, and Meca et al. (2003) revisit inventory cost games and the SOC-rule.
It is seen that the wider class of n-person Economic Production Quantity
(EPQ) inventory situations with shortages lead to exactly the same class of
cost games. Moreover, an alternative characterization of the SOC-rule is
provided there. Mosquera et al. (2005) introduce the property of immunity to
coalition manipulation and demonstrate that the SOC-rule is the unique
solution for inventory cost games which satisfies this property.

Toledo (2002) analyzes the class of inventory games that arises from
inventory problems with special sale prices. A collective of firms trying to
minimize its joint inventory cost by means of cooperation may receive a
special discount on set-up cost just in ordering. Reasons for such a price
reduction range from competitive price wars to attempted inventory reduction
by the supplier. Each firm has its own set-up cost which is invariant to the
order size. In this paper we assume that when an order is being placed, it
is revealed that the supplier makes a special offer for the next order.
Notice that the above condition makes sense from an economic point of view
since if one firm is a very good client then the supplier himself would
benefit by giving the client preferential treatment. Cooperation among firms
is given by sharing order process and warehouse facilities: firms in a
coalition make their orders jointly and store in the cheapest warehouse.
This cooperative situation generates the class of inventory games with non
discriminatory temporary discounts.

The organization of this paper is as follows. We start by introducing
definitions and notations in section \ref{preli}. In section \ref{Temp} we
give a complete description of the inventory problem with temporary
discounts (henceforth IPTD). A natural variant of this problem in which
several agents each one facing an IPTD decide to cooperate to increase
benefits (inventory situations with non discriminatory temporary discounts)
is addressed in section \ref{IGWT}. Then, for each one of these situations
the corresponding cooperative game, namely inventory game with non
discriminatory temporary discounts, is defined. This new class of games
motivates the study of a more general class of TU games, namely $p$-additive
games. In section \ref{padditive}, we introduce the latter and study its
properties. It contains the class of inventory games with non discriminatory
temporary discounts as well as the class of inventory cost games. Several
properties (as total balancedness, monotonicity, convexity and concavity)
are analyzed for this class of games. Finally, in section \ref{charac} the
modified SOC-rule (a kind of proportional rule) is proposed and study. It is
a core-allocation which can be reached through a pmas. Moreover, we provides
two different characterizations of the modified SOC-rule.

\section{\label{preli}Preliminaries}

A cooperative TU game is a pair $(N,w)$, where $N=\left\{ 1,2,...,n\right\} $
is the finite player set, $\mathcal{P}(N)$ is the set of all coalitions in $%
N,$ and $w:\mathcal{P}(N)\rightarrow \mathbb{R}$ the characteristic function
satisfying $w(\varnothing )=0.$ The subgame related to coalition $S,w_{S},$
is the restriction of mapping $w$ to the subcoalitions of $S.$ We denote by
small letter $s$ the cardinality of set $S $ , i.e. $card(S)=s,$ for all $%
S\subseteq N.$ We denote by $G$ the class of all TU games. A special family
of TU games is the family of unanimity games $\left\{ u_{S}\right\}
_{S\subseteq N}.$ The unanimity game associate to coalition $T\subseteq N$\
is defined by $u_{T}(S)=1$\ if $T\subseteq S$\ and $u_{T}(S)=0$\ otherwise.
An allocation vector will be $x\in \mathbb{R}^{n}$ and, for every coalition $%
S\subseteq N$ we shall write $x(S):=\sum_{i\in S}x_{i}$ the allocation to
coalition $S$ (where $x(\varnothing )=0).$

We say that a TU game is a cost game if its characteristic function
represents costs for all coalitions in $N;$\ it will be denoted by $(N,c).$\
On the other hand, a benefit game is a TU game whose characteristic function
represents benefits for all coalitions of $N;$\ now it will be denoted by $%
(N,v).$\ The reader may notice that a generic game will be denoted by $%
(N,w). $

Generally speaking, the core of a generic game $(N,w)$\ consists of all
those allocation vectors which are stable in the sense that none coalition
has incentive to leave the grand coalition; i.e. it is profitable for all
players in the coalition to take part of the grand coalition. Specifically,
the core of a cost game $(N,c)$ is given by $C(N,c)=\left\{ x\in \mathbb{R}%
^{n}\left/ x(N)=c(N)\text{ and }x(S)\leq c(S)\text{ for all }S\subset
N\right. \right\} ;$ and for a benefit game by $C(N,v)=\left\{ x\in \mathbb{R%
}^{n}\left/ x(N)=v(N)\text{ and }x(S)\geq v(S)\text{ for all }S\subset
N\right. \right\} .$ From now on allocation vectors belonging to the core
will be called core-allocations. A TU game $(N,w)$ has a nonempty core if
and only if it is balanced (see Bondareva, 1963 and Shapley, 1967). It is a
totally balanced game if the core of every subgame is nonempty.

A population monotonic allocation scheme (Sprumont, 1990), or pmas, for the
game $(N,w)$ is a collection of vectors $y^{S}\in \mathbb{R}^{s}$ for all $%
S\subseteq N,S\neq \varnothing $ such that $y^{S}(S)=w(S)$ for all $%
S\subseteq N,S\neq \varnothing .$ Moreover, for each player $i$\ in any
coalition $S$\ is more profitable $y_{i}^{T}$\ than $y_{i}^{S}$\ provided $%
S\subseteq T\subseteq N.$\ Logically this condition is $y_{i}^{S}\geq
y_{i}^{T}$\ for a cost game and $y_{i}^{S}\leq y_{i}^{T}$\ for a benefit game%
$.$ Note that if $\left( y^{S}\right) _{\varnothing \neq S\subseteq N}$ is a
pmas for $(N,w),$ then $y^{S}\in C(N,w_{S})$ for all $S\subseteq N,S\neq
\varnothing .$ Hence, a pmas can be seen as a refinement of the core. Every
TU game with pmas is totally balanced. A core-allocation for $(N,w),$ i.e. $%
x\in C(N,w)$, is reached through a pmas if there exits a pmas $\left(
y^{S}\right) _{\varnothing \neq S\subseteq N}$ \ for the game $(N,w)$ such
that $y_{i}^{N}=x_{i}$ for all $i\in N.$

A game is said to be monotone increasing when for all coalitions $S\subseteq
$ $T\subseteq N,w(S)\leq w(T)$ and monotone decreasing if $w(S)\geq w(T).$
We say that a game is superadditive when for all disjoint coalitions $S$ and
$T,$ $w(S\cup T)\geq w(S)+w(T)$ holds and subadditive if $w(S\cup T)\leq
w(S)+w(T)$ holds. A well-known class of balanced and subadditive
(superadditive) games is the class of concave (convex) games (Shapley,
1971). A TU game $(N,w)$ is concave (convex) if and only if $w(S\cup
\{i\})-w(S)\geq w(T\cup \{i\})-w(T)$ $\left( w(S\cup \{i\})-w(S)\leq w(T\cup
\{i\})-w(T)\right) $ for all player $i\in N$ and all pair of coalitions $%
S,T\subseteq N$ such that $S\subseteq T\subseteq N\backslash \{i\}.$

Another class of balanced and subadditive games is the class of
permutationally concave games (Granot and Huberman, 1982). Denote by $\Pi
(N) $ the set of all orders in $N.$ For any $\sigma \in \Pi (N)$\ and for
all $i\in N,\sigma (i)=j$ means that with respect to $\sigma $, player $i$
is in the $j$ $-$ $th$ position. Besides, $P_{i}^{\sigma }=\{j\in N/\sigma
(j)<\sigma (i)\}$ is the set of predecessors of $i$ with respect to $\sigma $
excluding $i$, and $\overline{P}_{i}^{\sigma }=\{j\in N/\sigma (j)\leq
\sigma (i)\}=P_{i}^{\sigma }\cup \{i\}$ is the set of predecessors of $i$
with respect to $\sigma $ including $i$. Define for all $\sigma \in \Pi
(N),P_{0}^{\sigma }=\varnothing .$ A cost game $(N,c)$ is permutationally
concave with respect to $\sigma \in \Pi (N)$ if and only if $c(\overline{P}%
_{i}^{\sigma }\cup R)-c(\overline{P}_{i}^{\sigma })\geq c(\overline{P}%
_{j}^{\sigma }\cup R)-c(\overline{P}_{j}^{\sigma })$ for all $i,j\in N\cup
\{0\}$ such that $\sigma (i)\leq \sigma (j)$ and all $R\subseteq N\setminus
\overline{P}_{j}^{\sigma }$. A game is permutationally concave if and only
if there exists an order $\sigma \in \Pi (N)$ such that the game is
permutationally concave with respect to $\sigma $.

The proportional rule with respect to $\lambda \in \mathbb{R}^{n}\backslash
\{0_{n}\},$ or $\lambda -$proportional rule, is a linear operator on the
class of all TU games and for a game $(N,w)$ is defined as $p(w)=\left[
\lambda /\lambda (N)\right] \cdot w(N).$

Inventory cost games were introduced in Meca et al. (2004) as models for
inventory situations. The player set $N$ consists of a group of firms
dealing with the ordering and holding of a certain commodity. In an
inventory cost game, a group of players minimize their total cost by placing
their orders together as one big order (paying a fix ordering cost $a$). To
coordinate the ordering policy of the firms, some minimum public information
is needed: the optimal number of orders for each player, i.e. $m_{i}$ for
all $i\in N.$ Then an inventory cost situation is given by the 3-tuple $%
\left\langle N,a,\{m_{i}\}_{i\in N}\right\rangle $ with $a>0$ and $m_{i}\geq
0,$ for all $i\in N.$ The corresponding inventory cost game $(N,c)$ is
defined as follows. For all non-empty coalitions $S\subseteq N,$

\begin{equation}
c(S):=2a\left( \sum_{i\in S}m_{i}^{2}\right) ^{1/2}.  \label{icg}
\end{equation}

We denote by $I$ the class of all inventory cost games. Meca et al. (2004)
shows that inventory cost games are concave and monotone. Moreover, the $%
c^{2}-$proportional rule with $c^{2}=\left( c(\left\{ i\right\} )^{2}\right)
_{i\in N},$ or SOC-rule, on inventory cost games is a core-allocation which
can be reached through a pmas for $(N,c).$ In addition, the SOC-rule is the
unique rule on the class of inventory cost games satisfying efficiency,
symmetry and monotonicity. Meca et al. (2003) revisit inventory cost games
and the SOC-rule. They prove that the wider class of $n$-person EPQ
inventory situations with shortages lead to exactly the same class of cost
games. Moreover, an alternative characterization of the SOC-rule, based on
an ad hoc additivity property, is provided there.

\section{\label{Temp}Temporary discounts on inventory problems}

It is a common practice for suppliers to offer special sale prices on orders
as an economic incentive to buyers to purchase in larger lot sizes. They may
temporarily discount the unit price of a product during a regular
replenishment cycle. Reasons for such a price reduction reach from
competitive prices wars to attempt inventory reduction. The rational
reaction to finding a product on sale during a regular replenishment is to
order additional units to take advantage of the short-live price reduction.
If a special order is arranged, then management must determine the optimum
order size and maximum shortage to place.

We consider a firm $i$ making orders of a certain product that it sells. The
fixed cost of an order is $a>0.$ The demand that he must fulfil is
deterministic and equal to $d_{i}$ units per time unit $(d_{i}\geq 0)$. The
cost of keeping in stock one unit of this product per time unit is $h_{i}$ $%
(h_{i}>0).$ Besides, firm $i$ considers the possibility of not fulfilling
all the demand in time, but allowing a shortage of the good. The cost of a
shortage of one unit of the good for one time unit is $s_{i}>0.$ When an
order is placed, after a deterministic and constant lead time (which can be
assumed to be zero, w.l.o.g.), firm $i$ receives the order gradually; more
precisely, $r_{i}$ units of the good are received per time unit. It is
assumed that $r_{i}>d_{i}$ (otherwise the model makes little sense). We call
$r_{i}$ the \textit{replacement rate} of agent $i$.

The inventory model we are dealing is the EPQ with shortages (see Meca et
al. (2003) for further details). The analysis of the EPQ model with
shortages and temporary discounts, that we will describe below, is inspired
by the EOQ model with special sale prices as introduced by Tersine (1994),
but it is new. Assume that when an order has been placed, it is revealed
that the supplier offers a special sale price for the next order. The
regular price of the product is $P$, but the next purchase can be made at $%
P-k$, where $k(\geq 0)$ is the unit price decrease. Subsequent to the
temporary sale, the price of the product will return to $P.$ Notice that the
order size and the maximum shortage that firm $i$ must choose prior and
after to the price decreases are those minimizing his average inventory cost
per time unit; i.e. $Q_{i}^{\ast }$ and $M_{i}^{\ast }$ where

\begin{equation*}
\begin{array}{cc}
Q_{i}^{\ast }=\sqrt{\frac{2ad_{i}}{h_{i}(1-\frac{d_{i}}{r_{i}})}\left( \frac{%
h_{i}+s_{i}}{s_{i}}\right) }, & M_{i}^{\ast }=\sqrt{\frac{2ad_{i}h_{i}}{%
s_{i}(h_{i}+s_{i})}\left( 1-\frac{d_{i}}{r_{i}}\right) }.%
\end{array}%
\end{equation*}

To obtain the optimal special order size and maximum shortage, it is
necessary to maximize the cost difference during the time period $\frac{Q_{i}%
}{d_{i}}$ with and without the special order.

The total cost during the period $\frac{Q_{i}}{d_{i}}(d_{i}>0)\footnote{%
It is assumed that $TC_{D}(Q_{i},M_{i})=0$ if $d_{i}=0.$}$, when a special
order is purchased at unit price $P-k$, is as follows:

\begin{equation}
TC_{D}(Q_{i},M_{i})=\underset{\text{order cost}}{\underbrace{a}}+\underset{%
\text{holding cost}}{\underbrace{\frac{h_{i}\left( Q_{i}\left( 1-\frac{d_{i}%
}{r_{i}}\right) -M_{i}\right) ^{2}}{2d_{i}\left( 1-\frac{d_{i}}{r_{i}}%
\right) }}}+\underset{\text{shortage cost}}{\underbrace{\frac{s_{i}M_{i}^{2}%
}{2d_{i}\left( 1-\frac{d_{i}}{r_{i}}\right) }}}+\underset{\text{purchase cost%
}}{\underbrace{\left( P-k\right) Q_{i}}}.
\end{equation}

If no special order is placed during $\frac{Q_{i}}{d_{i}},$ the total cost
when the first order is made at $P-k$ and all subsequent orders are made at $%
P$ is as follows:

\begin{eqnarray}
TC_{N}(Q_{i}) &=&\underset{\text{order cost}}{\underbrace{a\frac{Q_{i}}{%
Q_{i}^{\ast }}}}+\underset{\text{holding cost}}{\underbrace{\frac{Q_{i}}{%
d_{i}}\frac{h_{i}\left( Q_{i}^{\ast }\left( 1-\frac{d_{i}}{r_{i}}\right)
-M_{i}^{\ast }\right) ^{2}}{2Q_{i}^{\ast }\left( 1-\frac{d_{i}}{r_{i}}%
\right) }}}  \notag \\
&&+\underset{\text{shortage cost}}{\underbrace{\frac{Q_{i}}{d_{i}}\frac{%
s_{i}M_{i}^{\ast 2}}{2Q_{i}^{\ast }\left( 1-\frac{d_{i}}{r_{i}}\right) }}}+%
\underset{\text{purchase cost}}{\underbrace{\left( P-k\right) Q_{i}^{\ast
}+P\left( Q_{i}-Q_{i}^{\ast }\right) }}
\end{eqnarray}

To find the optimal one-time special order size $\left( \overline{Q}%
_{i}\right) $ and maximum shortage $\left( \overline{M}\right) _{i}$, the
difference in total cost must be maximized. So taking into account that the
special order cost saving is given by

\begin{eqnarray*}
TC(Q_{i},M_{i}) &=&TC_{N}(Q_{i})-TC_{D}(Q_{i},M_{i}) \\
&=&a\left( \frac{Q_{i}}{Q_{i}^{\ast }}-1\right) +\frac{h_{i}Q_{i}Q_{i}^{\ast
}}{2d_{i}}\frac{s_{i}}{h_{i}+s_{i}}\left( 1-\frac{d_{i}}{r_{i}}\right) -%
\frac{h_{i}\left( Q_{i}\left( 1-\frac{d_{i}}{r_{i}}\right) -M_{i}\right) ^{2}%
}{2d_{i}\left( 1-\frac{d_{i}}{r_{i}}\right) } \\
&&-\frac{s_{i}M_{i}^{2}}{2d_{i}\left( 1-\frac{d_{i}}{r_{i}}\right) }+k\left(
Q_{i}-Q_{i}^{\ast }\right) ,
\end{eqnarray*}%
it turns out that

\begin{equation*}
\begin{array}{l}
\overline{Q}_{i}=Q_{i}^{\ast }+\frac{k\cdot d_{i}}{h_{i}\left( 1-\frac{d_{i}%
}{r_{i}}\right) }\left( \frac{h_{i}+s_{i}}{s_{i}}\right) =\sqrt{\frac{2ad_{i}%
}{h_{i}\left( 1-\frac{d_{i}}{r_{i}}\right) }\frac{h_{i}+s_{i}}{s_{i}}}+\frac{%
k\cdot d_{i}}{h_{i}\left( 1-\frac{d_{i}}{r_{i}}\right) }\left( \frac{%
h_{i}+s_{i}}{s_{i}}\right) , \\
\overline{M}_{i}=\frac{h_{i}}{h_{i}+s_{i}}\overline{Q}_{i}\left( 1-\frac{%
d_{i}}{r_{i}}\right) .%
\end{array}%
\end{equation*}

The reader may notice that when the unit price discount is zero $\left(
k=0\right) ,$ the formulas for the optimum special order size and maximum
shortage reduce to the EPQ with shortages formulas and the cost saving is
zero $\left( TC(Q_{i}^{\ast },M_{i}^{\ast })=0\right) .$

It is easy to check that the optimum cost saving is

\begin{equation*}
TC(\overline{Q}_{i},\overline{M}_{i})=k^{2}\left( \frac{d_{i}}{2h_{i}\left(
1-\frac{d_{i}}{r_{i}}\right) }\right) \left( \frac{h_{i}+s_{i}}{s_{i}}%
\right) =k^{2}\frac{d_{i}^{2}}{4am_{i}^{2}},
\end{equation*}%
where $m_{i}=\frac{d_{i}}{Q_{i}^{\ast }}$ is the optimal number of orders
per unit of time if there is no special sale prices.

Since $TC(\overline{Q}_{i},\overline{M}_{i})$ is non negative, it is always
desirable to place a special order when a unit price discount is encountered
during a regular replenishment.

\section{\label{IGWT}Inventory games with non discriminatory temporary
discounts}

Once we have described inventory problems with temporary discounts (IPTD),
we address a natural variant of this problem in which several firms facing
each one a (IPTD) decide to cooperate in order to reduce costs. Here the
cooperation is driven by sharing order process and warehouse facilities.
Thus, if a group of firm agree on cooperation then they will make their
orders jointly and store in the cheapest warehouse.

Assume that the firms in $S\subseteq N$ decide to make their orders jointly
to save part of the order costs. We will consider situations in which there
is full disclosure of information. Each agent $i\in S$ reveals its demand $%
d_{i},$ holding cost $h_{i},$ shortage cost $s_{i},$ replacement rate\textit{%
\ }$r_{i}$, its individual optimal order size $\overline{Q}_{i}$ and maximum
shortage $\overline{M}_{i}$. In addition, if we assume there are no limits
to storage capacities, transport costs are equal to zero and deterministic
transport times, then we can consider coordination with regard to holding
cost. If a member of a coalition $S$ has a very low holding cost then this
coalition can reduce its cost by storing its inventory in the warehouse of
this member.

Following the same reasoning in Meca et al. (2004), it can be easily checked
that, in order to minimize the sum of the average inventory costs per time
unit, the agents must coordinate their orders so $\widehat{Q}_{i}/d_{i}=%
\widehat{Q}_{j}/d_{j}$ for all $i,j\in N,$ where $\widehat{Q}_{i}$ and $%
\widehat{Q}_{j}$ denoting the optimal order sizes for $i$ and $j$ if agents
in $S$ cooperate. Moreover, all goods will be stored in the warehouse of the
agent with the lowest holding cost. Define $h_{S}:=\min_{j\in S}\{h_{j}\}.$
Then the special order cost saving is given by

\begin{equation*}
TC(Q_{i},(M_{j})_{j\in S})=TC_{N}(Q_{i})-TC_{D}(Q_{i},(M_{j})_{j\in S}),
\end{equation*}%
where

\begin{eqnarray*}
TC_{D}(Q_{i},(M_{j})_{j\in S}) &=&a+\frac{Q_{i}}{d_{i}}\sum_{j\in S}\frac{%
h_{S}\left( Q_{j}\left( 1-\frac{d_{j}}{r_{j}}\right) -M_{j}\right) ^{2}}{%
2Q_{j}\left( 1-\frac{d_{j}}{r_{j}}\right) }+\frac{Q_{i}}{d_{i}}\sum_{j\in S}%
\frac{s_{j}M_{j}^{2}}{2Q_{j}\left( 1-\frac{d_{j}}{r_{j}}\right) } \\
&&+\left( P-k\right) \sum_{j\in S}Q_{j},
\end{eqnarray*}%
and

\begin{eqnarray*}
TC_{N}(Q_{i}) &=&a\frac{Q_{i}}{\widehat{Q}_{i}^{\ast }}+\frac{Q_{i}}{d_{i}}%
\sum_{j\in S}\frac{h_{S}\left( \widehat{Q}_{j}^{\ast }\left( 1-\frac{d_{j}}{%
r_{j}}\right) -\widehat{M}_{j}^{\ast }\right) ^{2}}{2\widehat{Q}_{j}^{\ast
}\left( 1-\frac{d_{j}}{r_{j}}\right) } \\
&&+\frac{Q_{i}}{d_{i}}\sum_{j\in S}\frac{s_{j}\widehat{M}_{j}^{\ast 2}}{2%
\widehat{Q}_{j}^{\ast }\left( 1-\frac{d_{j}}{r_{j}}\right) }+\left(
P-k\right) \sum_{j\in S}\widehat{Q}_{j}^{\ast }+P\left( \sum_{j\in S}\left(
Q_{j}-\widehat{Q}_{j}^{\ast }\right) \right)
\end{eqnarray*}%
with

\begin{equation*}
\begin{array}{l}
\widehat{Q}_{j}^{\ast }=\sqrt{\frac{2ad_{j}^{2}}{h_{S}\sum_{k\in S}d_{k}%
\frac{s_{k}}{h_{S}+s_{k}}\left( 1-\frac{d_{k}}{r_{k}}\right) }} \\
\widehat{M}_{j}^{\ast }=\widehat{Q}_{j}^{\ast }\frac{h_{S}\left( 1-\frac{%
d_{j}}{r_{j}}\right) }{h_{S}+s_{j}}%
\end{array}%
\end{equation*}%
for all $j\in S.$

Applying standard techniques of differential analysis it can be checked that
the values $(\hat{Q}_{i})_{i\in S}$ and $(\hat{M}_{i})_{i\in S}$ which
maximize $TC$ are given by:

\begin{equation*}
\begin{array}{l}
\hat{Q}_{i}=\widehat{Q}_{i}^{\ast }+k\cdot \frac{d_{i}\sum_{j\in S}\text{ }%
d_{j}}{h_{S}\sum_{j\in S}d_{j}\left( 1-\frac{d_{j}}{r_{j}}\right) \frac{s_{j}%
}{h_{S}+s_{j}}} \\
\hat{M}_{i}=\frac{h_{S}}{h_{S}+s_{i}}\hat{Q}_{i}\left( 1-\frac{d_{i}}{r_{i}}%
\right)%
\end{array}%
\end{equation*}%
for all $i\in S$. From this it follows that the maximal cost saving for
coalition $S$ equals
\begin{equation*}
TC(\hat{Q}_{i},(\hat{M}_{j})_{j\in S})=k^{2}\frac{\left( \sum_{j\in S}\text{
}d_{j}\right) ^{2}}{2h_{S}\sum_{j\in S}\text{ }d_{j}\left( 1-\frac{d_{j}}{%
r_{j}}\right) \frac{s_{j}}{h_{S}+s_{j}}}=k^{2}\frac{\left( \sum_{j\in S}%
\text{ }d_{j}\right) ^{2}}{4am_{S}^{2}}
\end{equation*}%
where $m_{S}=\frac{d_{i}}{\widehat{Q}_{i}^{\ast }}=\sqrt{\sum_{i\in
S}m_{i}^{2}}$ is the optimal number of orders per unit of time for coalition
$S$ if there is no special sale prices\footnote{%
Notice that $\frac{d_{i}}{\widehat{Q}_{i}^{\ast }}=\frac{d_{j}}{\widehat{Q}%
_{j}^{\ast }}$ for all $i,j\in S.$}.

The reader may notice that if a group of firms $S\subseteq N$ facing each
one a (IPTD) decide to cooperate by making their orders jointly and storing
in the cheapest warehouse, they will always be able to obtain a maximal cost
saving $TC(\hat{Q}_{i},(\hat{M}_{j})_{j\in S})\geq 0.$ \medskip

Next we consider the probability vector $\lambda =\left( \lambda \left(
S\right) \right) _{\emptyset \neq S\subseteq N},$ with $\lambda \left(
S\right) $ being the probability of coalition $S\subseteq N$ find a special
offer when ordering. Note that the above probability vector $\lambda $
should be designed by the supplier however he may want to do it. Then, for
each nonempty coalition $S\subseteq N$, the maximal average cost saving is
given by $\lambda \left( S\right) TC(\hat{Q}_{i},(\hat{M}_{j})_{j\in S}).$

Taking into account all above mentioned, a new cooperative situation can be
described by the tuple $\left\langle N,a,d,m,k,\lambda \right\rangle $ where
$N=\{1,...,n\}$ is the set of firms, $a>0$ is the ordering cost, $%
d=(d_{1},...,d_{n})$ is the vector of demands, $m=(m_{1},...,m_{n})$ is the
vector of individual optimal number of orders per unit of time if there is
no special sale price $(d_{i},m_{i}\geq 0$ for all $i\in N)$, $k$ is the
unit price discount $(k\geq 0$), and $\lambda $ is the probability vector $%
(0\leq \lambda \left( S\right) \leq 1,\forall S\subseteq N).$ This tuple
will be called \emph{inventory situation with temporary discounts.} \medskip

There exist a lot of real situations where the supplier gets benefits from
rewarding his customers according to the number of orders and the order
frequency just after finding special offers. Let's say, for instance,
pharmaceutical companies or computer manufacturers. The more they sell (in
terms of frequency and quantity) the better the balance sheet is. The
increase in liquidity and cash flow is also an added benefit when the
supplier needs to cover expenses, pay off debts in the short term, or even
if he want to increase its stock.

Assume that the supplier designs the probability vector $\lambda $ taking
into account the following criteria: customer order loyalty, customer order
frequency and liquidity, but does not exclude special sale prices for the
grand coalition $(\lambda (N)\neq 0).$ Formally, let $\left\langle
N,a,d,m,k,\lambda \right\rangle $ be an inventory situation with temporary
discounts. For every nonempty coalitions $S,T\subseteq N$

\begin{itemize}
\item[(i)] the \emph{order index between }$S$ \emph{and} $T$ is defined as
follows: $I_{m}(S,T):=m_{S}/m_{T},$ where $m_{R}$ denotes the optimal number
of orders per unit of time for coalition $R\subseteq N$ if there is no
special sale prices. The above index measures the degree of customer order
loyalty to the supplier by each coalition if there is no special sale
prices. Then if the supplier rewards the order loyalty of his clients, the
greater the order index between $S$ and $T$ is the benefit for $S$ should
increase more than for $T.$

\item[(ii)] the \emph{waiting index for }$S$ \emph{and} $T$ is defined by $%
I_{t}(S,T):=t_{S}/t_{T},$ where $t_{R}$ denotes the time between two
consecutive orders by coalition $R\subseteq N$ with special sale prices for
the first one. It measures the degree of customer order frequency to the
supplier by each coalition just after finding special sale prices. Again, if
the supplier rewards the order frequency of his clients, the greater the
waiting index for $S$ and $T$ is the greater the benefit for $T$ than for $S$
is.

\item[(iii)] the \emph{liquidity index between }$S$ \emph{and} $T$ is
defined as follows: $I_{l}(S,T):=l_{S}/l_{T},$ where $l_{R}$ denotes the
ratio between the order sizes by coalition $R\subseteq N$ with and without
special sale prices. This index measures the capacity of both coalitions to
provide immediate liquidity to the supplier if there is special sale prices.
Then if the supplier offers special sale prices in order to get liquidity,
the greater the liquidity index between $S$ and $T$ is the special price
policy treat more favorable to $S$ than $T$.
\end{itemize}

Therefore, the probability vector $\lambda $ based on customer order
loyalty, customer order frequency and liquidity criteria should satisfy the
following property: the ratio $\frac{\lambda (S)}{\lambda (N)}$ increases
with respect to order and liquidity indexes between $S$ and $N$ and
decreases with respect to the waiting index for $S$ and $N;$ i.e., there
exists $0<\alpha \leq 1$ such that for every nonempty coalition $S\subseteq
N,\lambda (S)/\lambda (N)=\alpha I_{m}(S,N)I_{l}(S,N)/I_{t}(S,N).$ From now
on $\left\langle N,a,d,m,k,\lambda \right\rangle $ with $\lambda $
satisfying the latter condition will be called \emph{inventory situation
with non discriminatory temporary discounts.}

The reader may notice that designing special price policies in such a way
gives to coalition $S$ only a concession: the one obtained from the customer
order loyalty or frequency as well as from the capacity to provide immediate
liquidity to the supplier, both measured by the above indexes. \medskip

Next Proposition shows that the ratio $\lambda (S)/\lambda (N)$ increases
with respect to the optimal number of orders per unit of time for coalition $%
S\subseteq N$ if there is no special sale prices, and decreases with respect
to the one for the grand coalition.

\begin{proposition}
\label{ND}Let $\left\langle N,a,d,m,k,\lambda \right\rangle $ be an
inventory situation with non discriminatory temporary discounts. There
always exists $0<\alpha \leq 1$ such that $\lambda (S)/\lambda (N)=\alpha
m_{S}^{2}/m_{N}^{2},$ for every nonempty coalition $S\subseteq N.$
\end{proposition}

\begin{proof}
The \emph{waiting index for }$S$ \emph{and} $N$ can be rewritten as%
\begin{equation*}
I_{t}(S,N)=\frac{\left( \frac{1}{m_{S}}+k\frac{\sum_{i\in S}d_{i}}{%
2am_{S}^{2}}\right) }{\left( \frac{1}{m_{N}}+k\frac{\sum_{i\in N}d_{i}}{%
2am_{N}^{2}}\right) }=\frac{m_{N}^{2}}{m_{S}^{2}}\left( \frac{%
2am_{S}+k\sum_{i\in S}d_{i}}{2am_{N}+k\sum_{i\in N}d_{i}}\right) ,
\end{equation*}%
and the \emph{liquidity index between }$S$ \emph{and} $N$ as%
\begin{eqnarray*}
I_{l}(S,N) &=&\frac{\left( \left. \sum_{i\in S}\hat{Q}_{i}\right/ \sum_{i\in
S}\widehat{Q}_{i}^{\ast }\right) }{\left( \left. \sum_{i\in N}\hat{Q}%
_{i}\right/ \sum_{i\in N}\widehat{Q}_{i}^{\ast }\right) }=\frac{\left.
\left( 2am_{S}+k\sum_{i\in S}d_{i}\right) \right/ 2am_{S}}{\left. \left(
2am_{N}+k\sum_{i\in N}d_{i}\right) \right/ 2am_{N}} \\
&=&\frac{m_{N}}{m_{S}}\left( \frac{2am_{S}+k\sum_{i\in S}d_{i}}{%
2am_{N}+k\sum_{i\in N}d_{i}}\right) .
\end{eqnarray*}

Then, there exists $\alpha \in (0,1]$ such that%
\begin{equation*}
\frac{\lambda (S)}{\lambda (N)}=\alpha \frac{I_{m}(S,N)I_{l}(S,N)}{I_{t}(S,N)%
}=\alpha \frac{m_{S}^{2}}{m_{N}^{2}}.
\end{equation*}
\end{proof}

Now taking into account Proposition \ref{ND} we are able to formalize the
inventory game corresponding to an inventory situation with temporary
discounts in the following way.

Given an inventory situation with non discriminatory temporary discounts $%
\left\langle N,a,d,m,k,\lambda \right\rangle ,$ the corresponding \emph{%
inventory game with non discriminatory temporary discounts }$\left(
N,v\right) $ is defined as follows: for all non-empty coalition $S\subseteq
N $

\begin{equation}
v(S):=K_{\lambda }\left( \sum_{j\in S}d_{j}\right) ^{2},  \label{ndtdg}
\end{equation}%
where $K_{\lambda }:=\frac{\lambda \left( N\right) \alpha k^{2}}{4am_{N}^{2}}%
\geq 0$.

We denote by $ID$ the class of all inventory games with non discriminatory
temporary discounts.

Comparing the classes of inventory games with non discriminatory temporary
discounts and inventory cost games, we find out a common property underlying
(\ref{icg}) and (\ref{ndtdg}); specifically the $\frac{1}{2}$-additivity
property for $\left( N,v\right) $ and the $2$-additivity property for $%
(N,c), $ i.e. $v(S)^{\frac{1}{2}}=\sum_{i\in S}v(\{i\})^{\frac{1}{2}}$ and $%
c(S)^{2}=\sum_{i\in S}c(\{i\})^{2}$ for all non-empty coalition $S\subseteq
N,$ respectively. Next we focus on a more general class of TU games which
contains the aforementioned classes.

\section{\label{padditive}$p$-additive games}

As we have just announced a new class of TU games is introduced in this
section. It is an extension of the classes of inventory games with non
discriminatory \ temporary discounts $ID$ (Toledo, 2002) as well as the
class of inventory cost games $I$ (Meca et al., 2003).

The class of $p$-additive games $A^{p}$ for every $p\in \mathbb{R}\backslash
\{0\},$ can be defined in the following way:

\begin{equation*}
A^{p}:=\left\{ (N,w)\in G\left\vert
\begin{array}{l}
w(S)\geq 0\text{ for all }S\subseteq N, \\
w(S)=0\text{ if }w(\{i\})=0\text{ for all }i\in S, \\
w(S)^{p}=\sum_{i\in S_{+}}w(\{i\})^{p}\text{ for all }S\subseteq N%
\end{array}%
\right. \right\} ,
\end{equation*}%
where $S_{+}:=\left\{ i\in S\left/ w(\{i\})>0\right. \right\} .$ Sometimes,
to avoid confusion we denote $S_{+}^{w}$\ to stress that we are focusing on
the $p$-additive game $(N,w).$\ Notice that the zero game $(N,w_{0})$\ is
also a $p$-additive game; i.e. $(N,w_{0})\in A^{p}$\ for all $p\in \mathbb{R}%
\backslash \{0\}.$\ For convenience we assume that $(N,w_{0})\in A^{2}.$

The reader may notice that every subgame of a $p$-additive game is a $p$%
-additive game as well; i.e. for every non-empty $S\subseteq N,\left(
S,w_{S}\right) \in A^{p}.$ Moreover, $A^{2}=I$ and $A^{\frac{1}{2}}=ID.$

In what follows the main properties for $p$-additive games are presented.

Next Theorem shows that $p$-additive games with $p$ non-negative $(p>0)$ are
monotone increasing. In addition, they are either convex or concave
depending on the value of the parameter $p>0.$

\begin{theorem}
Let $(N,w)\in A^{p}$ with $p>0.$ Then (i) $\left( N,w\right) $ is monotone
increasing; (ii) $\left( N,w\right) $ is convex if $p\leq 1,$ and concave if
$p\geq 1.$
\end{theorem}

\begin{proof}
\begin{enumerate}
\item[(i)] Take $S\subseteq T\subseteq N.$ Then $\sum_{j\in
S}w(\{j\})^{p}\leq \sum_{j\in T}w(\{j\})^{p}$ since $w(\{j\})^{p}\geq 0$ for
all $j\in N.$ Taking into account that function $f:\mathbb{R}%
_{++}\rightarrow \mathbb{R}_{++}$ such that $f(x)=x^{\frac{1}{p}}$ with $p>0$
is monotone increasing, we obtain that $w(S)\leq w(T).$ Hence we can
conclude that $\left( N,w\right) $ is monotone increasing.

\item[(ii)] Take $i\in N$ and $S\subseteq T\subseteq N\backslash \{i\}.$ Then%
\begin{equation*}
w(S\cup \{i\})-w(S)=\left( \sum\limits_{j\in S\cup \{i\}}w(\{j\})^{p}\right)
^{\frac{1}{p}}-\left( \sum\limits_{j\in S}w(\{j\})^{p}\right) ^{\frac{1}{p}}.
\end{equation*}

Taking into account that function $f:\mathbb{R}_{++}\rightarrow \mathbb{R}%
_{++}$ such that $f(x)=\left( x+A\right) ^{\frac{1}{p}}-x^{\frac{1}{p}}$
with $A\geq 0$ constant, is monotone increasing if $0<p\leq 1$ and monotone
decreasing if $p\geq 1$ we can conclude that $\left( N,w\right) $ is convex
if $p\leq 1$ and concave if $p\geq 1.$
\end{enumerate}
\end{proof}

Note that $p$-additive games with $p$ non-negative are totally balanced.
Moreover, the structure for their cores is well-known (see Shapley, 1971).

The following examples shows that the above properties are not hold, in
general, for $p$-additive games with $p$ negative $(p<0)$.

\begin{example}
\label{nega1}Consider $(\{1,2,3\},w)\in A^{-1}$ given by

\begin{equation*}
\begin{tabular}{c||ccccccc}
\hline
$S$ & $\{1\}$ & $\{2\}$ & $\{3\}$ & $\{1,2\}$ & $\{1,3\}$ & $\{2,3\}$ & $%
\{1,2,3\}$ \\ \hline
$w(S)$ & $1$ & $\frac{1}{2}$ & $\frac{1}{3}$ & $\frac{1}{3}$ & $\frac{1}{4}$
& $\frac{1}{5}$ & $\frac{1}{6}$ \\ \hline
\end{tabular}%
\end{equation*}

The reader may notice that the above game is monotone strictly decreasing
and subadditive but not concave. Moreover, the core for the above game is
nonempty since $\left( \frac{1}{6},0,0\right) \in C(N,w).$ In fact, it is
given by%
\begin{equation*}
C(N,w)=\{x\in \mathbb{R}^{3}/x(N)=\frac{1}{6};-\frac{1}{30}\leq x_{1}\leq
\frac{5}{12},-\frac{1}{12}\leq x_{2}\leq \frac{11}{30},-\frac{1}{6}\leq
x_{3}\leq \frac{17}{60}\}.
\end{equation*}
\end{example}

\begin{example}
\label{nega2}Consider $(\{1,2,3\},w)\in A^{-1}$ now given by

\begin{equation*}
\begin{tabular}{c||ccccccc}
\hline
$S$ & $\{1\}$ & $\{2\}$ & $\{3\}$ & $\{1,2\}$ & $\{1,3\}$ & $\{2,3\}$ & $%
\{1,2,3\}$ \\ \hline
$w(S)$ & $1$ & $0$ & $\frac{1}{2}$ & $1$ & $\frac{1}{3}$ & $\frac{1}{2}$ & $%
\frac{1}{3}$ \\ \hline
\end{tabular}%
\end{equation*}

The above game is concave, hence subadditive, but not monotone. The core is
again nonempty since $\left( 0,0,\frac{1}{3}\right) \in C(N,w).$ It is given
by%
\begin{equation*}
C(N,w)=\left\{ (x_{1},0,x_{3})\in \mathbb{R}^{3}/x_{1}+x_{3}=\frac{1}{3};-%
\frac{1}{6}\leq x_{1}\leq 1,-\frac{2}{3}\leq x_{3}\leq \frac{1}{2}\right\} .
\end{equation*}
\end{example}

We wonder if every $p$-additive game with $p$\ negative exhibits the above
subadditivity, monotonicity and balancedness properties. Next results give
satisfactory answers.

For all $p$-additive game $(N,w)$ and all coalition $S\subseteq N$, $%
w(S)=w(S_{+})$ holds since $w(S)^{p}=\sum_{j\in
S_{+}}w(\{j\})^{p}=w(S_{+})^{p}$. The following Proposition gives a
necessary and sufficient condition for $p$-additive games with $p<0$\ to be
monotone strictly decreasing.

\begin{proposition}
Let $(N,w)\in A^{p}$ with $p<0.$ Then $\left( N,w\right) $ is monotone
strictly decreasing if and only if $w(\{i\})>0$ for all $i\in N.$
\end{proposition}

\begin{proof}
(only if) Take $S\subset T\subseteq N.$ Then $\sum_{j\in
T}w(\{j\})^{p}>\sum_{j\in S}w(\{j\})^{p}$ since $w(\{j\})^{p}>0$ for all $%
j\in N.$ Now taking into account that function $f:\mathbb{R}_{++}\rightarrow
\mathbb{R}_{++}$ such that $f(x)=x^{\frac{1}{p}}$ with $p<0$ is monotone
strictly decreasing, we obtain that $w(T)<w(S).$ Hence we can conclude that $%
\left( N,w\right) $ is monotone strictly decreasing.

(if) For all $i\in N,w(\{i\})>w(N)\geq 0$ since $(N,w)$ is monotone strictly
decreasing and non-negative.
\end{proof}

It is immediate to check that monotonicity decreasing property is satisfied
only for all those coalitions consisting of players with non zero individual
values.

\begin{corollary}
Let $(N,w)\in A^{p}$ with $p<0.$ Then $\left( N,w\right) $ satisfies that $%
w(T_{+})\leq w(S_{+})$ for all $S\subseteq T\subseteq N$.
\end{corollary}

Next Theorem shows that $p$-additive games with $p<0$ are subadditive and
totally balanced.

\begin{theorem}
Every $(N,w)\in A^{p}$ with $p<0$ is subadditive and totally balanced.
\end{theorem}

\begin{proof}
(subadditive) Take $S\subseteq T\subseteq N$ such that $S\cap T=\varnothing $
We can distinguish two cases

\begin{itemize}
\item If $w(S\cup T)=0$ then $w(S)=0$ and $w(T)=0.$

\item If $w(S\cup T)>0$ then $w(S\cup T)=w(S_{+}\cup T_{+})\leq w(S_{+})\leq
w(S_{+})+w(T_{+})=w(S)+w(T).$
\end{itemize}

Hence we can conclude that $\left( N,w\right) $ is subadditive.

(totally balanced) It is enough to prove that every subgame $(S,w_{S})$ has
a non-empty Core.

Take $x\in \mathbb{R}^{s}$ the allocation given by $x_{i}=w(S),x_{j}=0$ for
all $j\neq i.$ We have just to prove that $\sum_{k\in T}x_{k}\leq w(T)$ for
all $T\subset S.$

Again two cases should be distinguish:

\begin{itemize}
\item $T\subseteq S\backslash \{i\}.$ Then $\sum_{k\in T}x_{k}=0\leq w(T)$
by definition of $(S,w_{S}).$

\item Take $T\subset S$ such that $i\in T.$ Then $\sum_{k\in
T}x_{k}=w(S)=w(S_{+})\leq w(T_{+})=w(T)$.
\end{itemize}
\end{proof}

The Proposition below shows that for every $p$-additive games with $p<0$\ to
be permutationally concave is equivalent to be concave, and both are
equivalent to the number of players with non zero individual values is never
greater than 2.

The following technical Lemma is needed to prove the aforementioned
Proposition.

\begin{lemma}
\label{lemita}Let $(N,w)\in A^{p}$ with $p<0.$ Let $i\in N$ and $S,T\subset
N $ such that $S\subset T\subseteq N\backslash \{i\}.$ Then
\begin{equation}
w(S\cup \{i\})-w(S)<w(T\cup \{i\})-w(T)  \label{con}
\end{equation}%
if and only if the following conditions are satisfied: (i) $\left\vert
S_{+}\right\vert \geq 1;$ (ii) $S_{+}\neq T_{+};$ (iii) $i\in N_{+}.$
\end{lemma}

\begin{proof}
(if) Suppose that any of the above three conditions does not satisfy. If (i)
does not satisfy, then condition (\ref{con}) is equivalent to $%
w(\{i\})<w(T\cup \{i\})-w(T)$, which is a contradiction since $(N,w)$ is
subadditive. In case that (ii) is not satisfied $w(S)=w(T)$ and $w(S\cup
\{i\})=w(T\cup \{i\}).$ Hence (\ref{con}) leads to a contradiction. Finally,
if $i\notin N_{+}$ then $w(S\cup \{i\})=w(S)$ and $w(T\cup \{i\})=w(T)$
which is also a contradiction.

(only if) Let us see that conditions (i), (ii) and (iii), implies (\ref{con}%
). For all coalition $S\subseteq N\backslash \{i\}$ satisfying (i) and for
any $p<0$

\begin{equation*}
w(S\cup \{i\})-w(S)=\left( \sum\limits_{j\in \left( S\cup \{i\}\right)
_{+}}w(\{j\})^{p}\right) ^{\frac{1}{p}}-\left( \sum\limits_{j\in
S_{+}}w(\{j\})^{p}\right) ^{\frac{1}{p}}.
\end{equation*}

Consider the function $f:\mathbb{R}_{++}\rightarrow \mathbb{R}_{++}$ such
that $f(x)=\left( x+A\right) ^{\frac{1}{p}}-x^{\frac{1}{p}}$ with $A\geq 0$
constant, is monotone strictly increasing if $p<0$ and $A>0.$ Therefore (ii)
implies that $\sum_{j\in S_{+}}w(\{j\})^{p}<\sum_{j\in T_{+}}w(\{j\})^{p}$
and by (iii) $A>0$. Hence (\ref{con}) holds.
\end{proof}

\begin{proposition}
Let $(N,w)\in A^{p}$ with $p<0.$ The following conditions are equivalent:
(i) $\left( N,w\right) $ is permutationally concave; (ii) $\left\vert
N_{+}\right\vert \leq 2;$ (iii) $\left( N,w\right) $ is concave$.$
\end{proposition}

\begin{proof}
$(i)\Longrightarrow (ii)$ Take $(N,w)\in A^{p}$ with $p<0$ to be
permutationally concave. If $\left\vert N_{+}\right\vert >2,$ let $%
i_{1}^{\sigma },i_{2}^{\sigma },i_{3}^{\sigma }\in N$ such that $%
w(i_{1}^{\sigma }),w(i_{2}^{\sigma }),w(i_{3}^{\sigma })>0$ and $\sigma
(i_{1}^{\sigma })<\sigma (i_{2}^{\sigma })<\sigma (i_{3}^{\sigma })$ for
each $\sigma \in \Pi (N)$. Then by Lemma \ref{lemita}%
\begin{equation*}
w\left( \overline{P_{i_{1}^{\sigma }}^{\sigma }}\cup \{i_{3}^{\sigma
}\}\right) -w\left( \overline{P_{i_{1}^{\sigma }}^{\sigma }}\right) <w\left(
\overline{P_{i_{2}^{\sigma }}^{\sigma }}\cup \{i_{3}^{\sigma }\}\right)
-w\left( \overline{P_{i_{2}^{\sigma }}^{\sigma }}\right) ,
\end{equation*}%
which is a contradiction.

$(ii)\Longrightarrow (iii)$ Suppose that $\left\vert N_{+}\right\vert \leq
2. $ We have to prove that%
\begin{equation}
w(S\cup \{i\})-w(S)\geq w(T\cup \{i\})-w(T)  \label{con1}
\end{equation}%
for all $i\in N$ and $S,T\subset N$ such that $S\subset T\subseteq
N\backslash \{i\}.$ If $w(\{i\})=0$ then $w(S\cup \{i\})=w(S)$ and $w(T\cup
\{i\})=w(T).$ If $w(\{i\})>0$ we consider the following cases:

\begin{enumerate}
\item $w(\{j\})>0$ for any $j\in S$. Then, $j\in T,w(S)=w(T)$ and $w(S\cup
\{i\})=w(T\cup \{i\}).$

\item $w(\{j\})>0$ for any $j\notin S.$ Then,

\begin{enumerate}
\item[2.1.] If $j\in T,w(S)=0,w(S\cup \{i\})=w(\{i\}),w(T\cup
\{i\})=w(\{i,j\})$ and $w(T)=w(\{j\})$. Hence (\ref{con1}) is equivalent to $%
w(\{i\})+w(\{j\})\geq w(\{i,j\})$ which is true since $(N,w)$ is subadditive.

\item[2.2.] If $j\notin T$, then $w(S)=w(T)=0$ and $w(S\cup \{i\})=w(T\cup
\{i\})=w(\{i\}).$
\end{enumerate}
\end{enumerate}

In all cases (\ref{con1}) is verified. Then $\left( N,w\right) $ is concave.

$(iii)\Longrightarrow (i)$ All concave games are permutationally concave
game.
\end{proof}

The following table summarizes all results we have obtained for $p$-additive
games:

\begin{equation*}
\begin{tabular}{|c|c|c|}
\hline
$p<0$ & \multicolumn{2}{|c|}{$p>0$} \\ \cline{2-3}
& $0<p\leq 1$ & $p\geq 1$ \\ \hline\hline
$%
\begin{array}{c}
\text{Permutationally concave} \\
\Longleftrightarrow \text{Concave}\Longleftrightarrow \left\vert
N_{+}\right\vert \leq 2%
\end{array}%
$ & Convex & Concave \\ \hline
Subadditive & Superadditive & Subadditive \\ \hline\hline
$%
\begin{array}{c}
\text{Monotone strictly decreasing} \\
\Longleftrightarrow w(\{i\})>0\text{ for all }i\in N%
\end{array}%
$ & \multicolumn{2}{|c|}{Monotone Increasing} \\ \hline\hline
\multicolumn{3}{|c|}{Totally balanced} \\ \hline
\end{tabular}%
\end{equation*}

The reader may notice that the interpretation of $p$-additive games could
change depending on the value of parameter $p.$ We mean that $p$-additive
games with $0<p\leq 1$ can be seen as benefit games (for instance $A^{\frac{1%
}{2}}=ID$ the class of inventory games with temporary discounts). However,
those $p$-additive games with $p\geq 1$ or $p<0$ should be noted as cost
games (for instance $A^{2}=I$ the class of inventory cost games).

\section{\label{charac}Modified SOC-rule}

Next goal is to find a core-allocation for $p$-additive games which is easy
to calculate and satisfies good properties. An appealing one point solution
concept for these games is the \emph{modified SOC-rule. }We will define the
modified SOC-rule for a non-zero game $(N,w)$ as the rule $\sigma ^{p}(N,w)$
that divides the grand coalition value $w(N)$ proportionally to the
individual values to the $p$. This implies that player $i\in N$ receives

\begin{equation*}
\sigma _{i}^{p}(N,w)=\frac{w(\{i\})^{p}}{\sum_{j\in N}w(\{j\})^{p}}w(N)=%
\frac{w(\{i\})^{p}}{w(N)^{p-1}}=w(\{i\})^{p}\left( \sum_{j\in
N}w(\{j\})^{p}\right) ^{\frac{1}{p}-1}.
\end{equation*}

For the zero game $(N,w_{0})$ each player receives nothing; i.e. $\sigma
_{i}^{p}(N,w_{0})=0$ for all player $i\in N.$

The reader may notice that the above rule coincides with the SOC-rule (Meca
et al., 2003) on the class of inventory cost games $A^{2}$.

Next we will see that this rule has some nice properties.

First, for all $p$-additive games $(N,w)$ it holds that $\sigma ^{p}(N,w)$
is a core-allocation. This is easy to see. It is enough to prove that for
all non-empty coalition $S\subset N,\sum_{i\in S}\sigma _{i}^{p}(N,w)\leq
w(S),$ if $p\geq 1$ or $p<0;\sum_{i\in S}\sigma _{i}^{p}(N,w)\geq w(S),$ if $%
0<p\leq 1.$

Taking into account that function $g:\mathbb{R}_{++}\rightarrow \mathbb{R}%
_{++}$ defined by $g(x)=x^{\frac{1}{p}-1}$ is monotone increasing if and
only if $0<p\leq 1$ and monotone decreasing if and only if $p\geq 1$ or $%
p<0, $ it holds that

\begin{eqnarray*}
\sum_{i\in S}\sigma _{i}^{p}(N,w) &=&\sum_{i\in S}w(\{i\})^{p}\left(
\sum_{j\in N}w(\{j\})^{p}\right) ^{\frac{1}{p}-1} \\
&\geq &\sum_{i\in S}w(\{i\})^{p}\left( \sum_{j\in S}w(\{j\})^{p}\right) ^{%
\frac{1}{p}-1}=\left( \sum_{i\in S}w(\{i\})^{p}\right) ^{\frac{1}{p}}=w(S),
\end{eqnarray*}%
if $0<p\leq 1,$ and

\begin{eqnarray*}
\sum_{i\in S}\sigma _{i}^{p}(N,w) &=&\sum_{i\in S}w(\{i\})^{p}\left(
\sum_{j\in N}w(\{j\})^{p}\right) ^{\frac{1}{p}-1} \\
&\leq &\sum_{i\in S}w(\{i\})^{p}\left( \sum_{i\in S}w(\{i\})^{p}\right) ^{%
\frac{1}{p}-1}=\left( \sum_{i\in S}w(\{i\})^{p}\right) ^{\frac{1}{p}}=w(S),
\end{eqnarray*}%
if $p\geq 1$ or $p<0$.

Second, this proportional rule can be reached through a pmas. Define for all
$i\in S,S\subseteq N$ and $S\neq \varnothing ,$

\begin{equation*}
y_{i}^{S}:=w(\{i\})^{p}\left( \sum_{j\in S}w(\{j\})^{p}\right) ^{\frac{1}{p}%
-1}.
\end{equation*}

Then for all non-empty coalition $S\subset N,\sum_{i\in S}y_{i}^{S}=w(S),$
and for all non-empty coalitions $S,T\subseteq N$ such that $S\subseteq T$
and for all $i\in S,y_{i}^{S}\geq y_{i}^{T},$ if $p\geq 1$ or $%
p<0;y_{i}^{S}\leq y_{i}^{T},$ if $0<p\leq 1.$ Finally, we see that $%
y_{i}^{N}=\sigma _{i}^{p}(N,w)$ for all $i\in N.$ Hence the rule $\sigma
^{p}(N,w)$ can be reached through the pmas $y.$

Note that for example \ref{nega1} the modified SOC-rule is $\sigma
^{-1}(N,w)=\left( \frac{1}{36},\frac{2}{36},\frac{3}{36}\right) $ and a pmas
through which is reached is given by

\begin{equation*}
\begin{tabular}{c||ccccccc}
\hline
$S$ & $\{1\}$ & $\{2\}$ & $\{3\}$ & $\{1,2\}$ & $\{1,3\}$ & $\{2,3\}$ & $%
\{1,2,3\}$ \\ \hline
$y^{S}$ & $1$ & $\frac{1}{2}$ & $\frac{1}{3}$ & $\left( \frac{1}{9},\frac{2}{%
9}\right) $ & $\left( \frac{1}{16},\frac{3}{16}\right) $ & $\left( \frac{2}{%
25},\frac{3}{25}\right) $ & $\left( \frac{1}{36},\frac{2}{36},\frac{3}{36}%
\right) $ \\ \hline
\end{tabular}%
\end{equation*}

For example \ref{nega2} the modified SOC-rule is $\sigma ^{-1}(N,w)=\left(
\frac{1}{9},0,\frac{2}{9}\right) $ and the corresponding pmas is now given by

\begin{equation*}
\begin{tabular}{c||ccccccc}
\hline
$S$ & $\{1\}$ & $\{2\}$ & $\{3\}$ & $\{1,2\}$ & $\{1,3\}$ & $\{2,3\}$ & $%
\{1,2,3\}$ \\ \hline
$y^{S}$ & $1$ & $0$ & $\frac{1}{2}$ & $\left( 1,0\right) $ & $\left( \frac{1%
}{9},\frac{2}{9}\right) $ & $\left( 0,\frac{1}{2}\right) $ & $\left( \frac{1%
}{9},0,\frac{2}{9}\right) $ \\ \hline
\end{tabular}%
\end{equation*}

We complete the study of the modified SOC-rule by presenting two different
characterizations for it. The first one is based on a kind of additivity
property (the so called $p$-Transfer) ad hoc for the class of $p$-additive
games, which is a straightforward generalization of the Transfer property
introduced by Meca et al. (2003). The second property, $p$-Monotonicity, it
is also inspired by the Monotonicity property used in Meca et al. (2004).

Let us start defining the $p$-sum on the class of $p$-additive games. Let $%
(N,w),$ $(N,w^{\prime })\in A^{p}.$ The $p$-sum of both games is: $\left(
w\oplus w^{\prime }\right) (S)=\left( w(S)^{p}+w^{\prime }(S)^{p}\right) ^{%
\frac{1}{p}}$ for all $S\subseteq N.$

It is easy to check that the class of $p$-additive games is closed for the $%
p $-sum (the $p$-sum of two $p$-additive games is a $p$-additive game); i.e
for all $\left( N,w\right) ,\left( N,w^{\prime }\right) \in A^{p}$ it holds $%
\left( N,w\oplus w^{\prime }\right) \in A^{p}.$ Note that when $p=1$ the $p$%
-sum is the usual sum in $\mathbb{R}.$

Consider $\left\{ u_{S}\right\} _{S\subseteq N}$ the family of unanimity
games with player set $N$. It is easy to check that every $u_{\{i\}}$ is a $%
p $-additive game with $i\in N.$ However, $u_{T}\notin A^{p}$ for $%
T\subseteq N $ with $\left\vert T\right\vert \geq 2$ since $u_{T}(T)\neq
\sum_{i\in T_{+}}u_{T}(\{i\})^{p}=0$.

Next step is to prove that any $p$-additive game can be expressed as a $p$%
-sum combination of unanimity games; i.e. the class of $p$-additive games is
generated by the set of unanimity games with a player only.

\begin{proposition}
\label{sumatorio}For all $(N,w)\in A^{p},$ a unique collection of
nonnegative scalars $\left\{ \alpha _{i}\right\} _{i\in N}$ exists such that
$w=\oplus _{i\in N}\alpha _{i}u_{\{i\}}.$
\end{proposition}

\begin{proof}
It is a straightforward generalization of the proof given in Meca et al.
(2003) for Proposition 1.
\end{proof}

Now, we introduce some properties that will allow us characterizing the
modified SOC-rule on the class of $p$-additive games. A solution $\varphi $\
on $p$-additive games is a map $\varphi :A^{p}\rightarrow \mathbb{R}^{N}.$
Then $\varphi (w)=\left( \varphi _{i}(w)\right) _{i\in N}$ where $\varphi
_{i}(w)$\ denotes the benefit to player $i\in N$\ according to this
allocation in the game $(N,w)\in A^{p}.$

Let $(N,w),(N,w^{\prime })$\ be $p$-additive games and $\varphi $ a solution
for them. We consider the following properties:

\begin{description}
\item[(EF)] Efficiency. $\sum_{i\in N}\varphi _{i}(w)=w(N).$

\item[(NP)] Null player. For all player $i\in N$ such that $w(\{i\})=0$ then
$\varphi _{i}(w)=0$.
\end{description}

Next two properties were proposed in Toledo (2002).

\begin{description}
\item[(PT)] $p$-Transfer. For all $i\in N$%
\begin{equation*}
\left( w\oplus w^{\prime }\right) (N)^{p-1}\varphi _{i}\left( w\oplus
w^{\prime }\right) =w(N)^{p-1}\varphi _{i}\left( w\right) +w^{\prime
}(N)^{p-1}\varphi _{i}\left( w^{\prime }\right) .
\end{equation*}
\end{description}

The reader may notice that the above property is a kind of transference from
the operation $p$-sum to the usual sum in $\mathbb{R}.$ A solution
satisfying $p$-Transfer gives to every player in the $p$-sum game the sum of
the solution values corresponding to each game, where all of these values
are pondered by the grand coalition value to the $p-1$.

\begin{description}
\item[(PMO)] $p$-Monotonicity. For all $i\in N$%
\begin{equation*}
w(\{i\})\geq w^{\prime }(\{i\})\Longrightarrow w(N)^{p-1}\varphi _{i}\left(
w\right) \geq w^{\prime }(N)^{p-1}\varphi _{i}\left( w^{\prime }\right)
\end{equation*}
\end{description}

Note that the above property is a generalization of the monotonicity
property used in Meca et al. (2004) to characterize the SOC-rule on the
class of inventory cost games. Moreover, (PT) and (PMO) are well defined for
all $p$-additive games since the zero game $(N,w_{0})\in A^{2}$ and then, $%
w_{0}(N)^{2-1}=0.$

Next Theorem states that there exists a unique solution on $p$-additive
games satisfying \textit{efficiency}, \textit{null player }and\textit{\ }$%
\mathit{p}$\textit{-Transfer }properties.

\begin{theorem}
\label{charac1}There exists a unique allocation on the class of $p$-additive
games satisfying (EF), (NP) and (PT). It is the modified SOC-rule.
\end{theorem}

\begin{proof}
It is a straightforward consequence of Theorem 1 in Meca et al. (2003).
\end{proof}

Last Theorem states that there also exists a unique solution on $p$-additive
games satisfying \textit{efficiency}, \textit{null player }and\textit{\ }$p$%
\textit{-Monotonicity }properties.

\begin{theorem}
\label{charac2}There exists a unique allocation on the class of $p$-additive
games satisfying (EF), (NP) and (PMO). It is the modified SOC-rule.
\end{theorem}

\begin{proof}
It is clear that the modified SOC-rule also satisfies (PMO). To prove the
converse we will use an induction argument on the cardinal of the set $%
N_{+}. $ Take a solution $\varphi $ on $p$-additive games that satisfies
(EF), (NP) and (PMO).

If $\left\vert N_{+}^{w}\right\vert =1$ then there exists a unique $j\in N$
such that $w(\{j\})>0$. Hence $\varphi (w)=\sigma ^{p}(N,w)$ since both of
them satisfy (EF) and (NP). Suppose that $\varphi (w)=\sigma ^{p}(N,w)$ for
all game $(N,w)\in A^{p}$ such that $1\leq \left\vert N_{+}^{w}\right\vert
\leq n-1.$

Let $(N,w^{\prime })\in A^{p}$ with $1\leq \left\vert N_{+}^{w^{\prime
}}\right\vert \leq n.$ Then, by Proposition \ref{sumatorio}%
\begin{equation*}
w^{\prime }=\oplus _{i\in N_{+}^{w^{\prime }}}w^{\prime
}(\{i\})u_{\{i\}}=\left( \oplus _{i\in N_{+}^{w^{\prime }}\setminus
\{k\}}w^{\prime }(\{i\})u_{\{i\}}\right) \oplus w^{\prime }(\{k\})u_{\{k\}}.
\end{equation*}

If we denote by $w$ the game $\oplus _{i\in N_{+}^{w^{\prime }}\setminus
\{k\}}w^{\prime }(\{i\})u_{\{i\}},$ it is easy to check that $%
w(\{i\})=w^{\prime }(\{i\})$ for all $i\in N\setminus \{k\}.$ By (PMO), $%
w(N)^{p-1}\varphi _{i}(w)=w^{\prime }(N)^{p-1}\varphi _{i}(w^{\prime }),$
for all $i\in N\setminus \{k\}.$ Now taking into account the induction
argument and $w^{\prime }(N)\neq 0$%
\begin{eqnarray*}
\varphi _{i}(w^{\prime }) &=&w^{\prime }(N)^{1-p}w(N)^{p-1}\varphi
_{i}(w)=w^{\prime }(N)^{1-p}w(N)^{p-1}\sigma _{i}^{p}(N,w) \\
&=&w^{\prime }(N)^{1-p}w(\{i\})=w^{\prime }(N)^{1-p}w^{\prime
}(\{i\})=\sigma _{i}^{p}(N,w^{\prime }),
\end{eqnarray*}%
for all $i\in N\setminus \{k\}.$ Finally, by (EF) $\varphi _{k}(w^{\prime
})=\sigma _{k}^{p}(N,w^{\prime }).$ By the other hand, if $\left\vert
N_{+}^{w}\right\vert =0,\varphi (w)=\sigma ^{p}(N,w)=0$ by (NP).
\end{proof}

The following examples show that (EF), (NP) and (PT) properties are
logically independent for Theorem \ref{charac1}.

\begin{example}
Consider $\varphi $ on $A^{p}$ defined by $\varphi _{i}(w)=\frac{\beta
w(\{i\})^{p}}{w(N)^{p}}w(N),$ for all player $i\in N,$ where $\beta \in
\mathbb{R}\backslash \{1\}.$ {$\varphi (w)$ satisfies (NP) and (PT) but not
(EF).}
\end{example}

\begin{example}
Take $\varphi $ on $A^{p}$ given by $\varphi _{i}(w)=\frac{w(N)}{n},$ for
all player $i\in N$. {$\varphi (w)$ satisfies (EF) and (PT) but not (NP).}
\end{example}

\begin{example}
{Shapley value (Shapley, 1953) satisfies (EF) and (NP) but not (PT)}.
\end{example}

Finally, to conclude this section, the three examples below show that (EF),
(NP) and (PMO) properties are logically independent for Theorem \ref{charac2}

\begin{example}
Consider $\varphi $ on $A^{p}$ defined by $\varphi
_{i}(w)=w(\{i\})w(N)^{1-p},$ for all player $i\in N.$ {$\varphi (w)$
satisfies (NP) and (PMO) but not (EF).}
\end{example}

\begin{example}
Take $\varphi $ on $A^{p}$ given by $\varphi _{i}(w)=\frac{w(\{i\})}{%
w(N)^{p-1}},$ for all player $i\in N$ and for all $(N,w)\in A^{p}$ such that
$\left\vert N_{+}\right\vert \geq 1$. By the other hand $\varphi
_{i}(w_{0})=1-n$ if $i=1$ and $\varphi _{i}(w_{0})=1$ otherwise. {$\varphi
(w)$ satisfies (EF) and (PMO) but not (NP).}
\end{example}

\begin{example}
Let $\varphi $ be a solution on $A^{p}$ defined by $\varphi _{i}(w)=\frac{%
w(N_{+})}{\left\vert N_{+}^{w}\right\vert }$ if $i\in N_{+}$ and $\varphi
_{i}(w)=0$ otherwise. {$\varphi (w)$ satisfies (EF) and (NP) but not (PMO).}
\end{example}

\newpage

\section*{References}

\noindent Bondareva ON (1963) Some applications of linear programming
methods to the theory of cooperative games. Problemy Kibernety 10:119-139.
In Russian.\newline

\noindent Granot DG and Huberman G (1982) The relation between convex games
and minimal cost spanning tree games: a case for permutationally convex
games. SIAM Journal of Algebraic and Discrete Methods 3:288-292. \newline

\noindent Meca A, Timmer J, Garc\'{\i}a-Jurado I and Borm PEM (2004)
Inventory Games. European Journal of Operational Research 156:127-139.%
\newline

\noindent Meca A, Garc\'{\i}a-Jurado I and Borm PEM (2003) Cooperation and
competition in Inventory Games. Mathematics Methods of Operation Research
57:481-493.\newline

\noindent Mosquera MA, Garc\'{\i}a-Jurado, I and Fiestras-Janeiro MG (2005)
A note on coalitional manipulation and centralized inventory management.
Annals of Operations Research (to appear).\newline

\noindent Shapley LS (1953) A Value for n-Person Games. In: Kuhn H, Tucker
AW (eds.) Contributions to the Theory of Games II. Princeton University
Press, pp. 307-317.\newline

\noindent Shapley LS (1967)\ On Balanced Sets and Cores. Naval Research
Logistics Quartely 14:453-460.\newline

\noindent Shapley LS (1971) Cores of Convex Games. International Journal of
Game Theory 1:11-26.\newline

\noindent Sprumont Y (1990) Population Monotonic Allocation Schemes for
Cooperative Games with Transferable Utility. Games and Economic Behavior
2:378-394.\newline

\noindent Tersine RJ (1994) Principles of Inventory and Material Management.
Amsterdam: Elsevier North Holland.\newline

\noindent Toledo A (2002) Problemas de Inventario con Descuento desde la
perspectiva de la Teor\'{\i}a de Juegos. Ph.D. thesis. Universidad Miguel
Hern\'{a}ndez de Elche.\newline


\end{document}